\documentclass[english]{article}
\usepackage{lmodern}
\usepackage[]{fontenc}
\usepackage[latin9]{inputenc}
\usepackage{color}
\usepackage{babel}

\usepackage{verbatim}
\usepackage{float}
\usepackage{units}
\usepackage{amsmath}
\usepackage{graphicx}
\usepackage{amssymb}
\usepackage[unicode=true, 
 bookmarks=true,bookmarksnumbered=true,bookmarksopen=false,
 breaklinks=false,pdfborder={0 0 0},backref=false,colorlinks=true]
 {hyperref}
\hypersetup{pdftitle={Quantum money from knots},
 pdfauthor={Edward Farhi, David Gosset, Avinatan Hassidim, Andrew Lutomirski, and Peter Shor}}
 
\makeatletter

\floatstyle{ruled}
\newfloat{algorithm}{tbp}{loa}
\floatname{algorithm}{Algorithm}

\usepackage{tikz}

\makeatother

\begin{document}

\title{Quantum money from knots}

\author{Edward Farhi, David Gosset, Avinatan Hassidim,\\
Andrew Lutomirski, and Peter Shor}

\date{April 28, 2010}
\maketitle
\begin{abstract}
Quantum money is a cryptographic protocol in which a mint can produce
a quantum state, no one else can copy the state, and anyone (with
a quantum computer) can verify that the state came from the mint.
We present a concrete quantum money scheme based on superpositions
of diagrams that encode oriented links with the same Alexander polynomial.
We expect our scheme to be secure against computationally bounded
adversaries.
\end{abstract}

\def\griddims(#1,#2){\def\negclip{\clip (-0.1,-0.1) rectangle (#1+0.1,#2+0.1) }}
\def\crossgap(#1,#2){ (#1-0.1,#2-0.1) rectangle (#1+0.1,#2+0.1) }
\long\def\clipseg#1#2{\begin{scope}[even odd rule] \negclip #1;#2 \end{scope}}


\def\hseg(#1,#2,#3){\draw (#1+0.1,#3) -- (#2-0.1,#3);}
\def\vseg(#1,#2,#3){\draw (#1,#2+0.1) -- (#1,#3-0.1);}

\def\larrow(#1,#2){\draw[->] (#1-0.45,#2) -- (#1-0.50,#2);}
\def\rarrow(#1,#2){\draw[->] (#1+0.45,#2) -- (#1+0.50,#2);}
\def\uarrow(#1,#2){\draw[->] (#1,#2+0.45) -- (#1,#2+0.50);}
\def\darrow(#1,#2){\draw[->] (#1,#2-0.45) -- (#1,#2-0.50);}

\def\mark_common(#1,#2){\fill[color=white] (#1-0.3,#2-0.3) rectangle (#1+0.3,#2+0.3);}
\def\redmark(#1,#2){\mark_common(#1,#2) \draw[line width=0.1em,line cap=round] (#1-0.2,#2-0.2) -- (#1+0.2,#2+0.2) (#1-0.2,#2+0.2) -- (#1+0.2,#2-0.2);}
\def\bluemark(#1,#2){\mark_common(#1,#2) \draw[line width=0.1em] (#1,#2) circle (0.2);}

\def\hsegfr(#1,#2,#3){\shade[left color=black,right color=white] (#1+0.1,#3-0.02) rectangle (#2-0.1,#3+0.02);}
\def\hsegfl(#1,#2,#3){\shade[left color=white,right color=black] (#1+0.1,#3-0.02) rectangle (#2-0.1,#3+0.02);}
\def\vsegfd(#1,#2,#3){\shade[top color=black,bottom color=white] (#1-0.02,#2+0.1) rectangle (#1+0.02,#3-0.1);}
\def\vsegfu(#1,#2,#3){\shade[top color=white,bottom color=black] (#1-0.02,#2+0.1) rectangle (#1+0.02,#3-0.1);}

\def\startmarks{\begin{scope}[shift={(0.5,0.5)},ultra thick]}
\def\endmarks{\end{scope}}

\def\xlabel(#1,#2){\draw[anchor=north] node at (#1+0.5,0) {\vphantom{$1$} #2};}  
\def\ylabel(#1,#2){\draw[anchor=east] node at (0,#1+0.5,0) {#2};}

\def\clipgrid(#1,#2,#3,#4){\clip (#1+0.1,#2+0.1) rectangle (#3-0.1,#4-0.1);}
\def\iclipgrid(#1,#2,#3,#4){\negclip (#1-0.04,#2-0.04) rectangle (#3+0.04,#4+0.04);}

\section{Introduction}

\begin{figure}[h]
\begin{centering}
\includegraphics{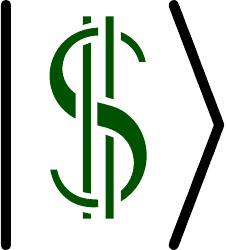}
\par\end{centering}

\caption{Quantum money from knots\label{fig:dollar-sign}}

\end{figure}

In this paper, we present quantum money based on knots (see Figure~\ref{fig:dollar-sign}).
The purported security of our quantum money scheme is based on the
assumption that given two different looking but equivalent knots,
it is difficult to explicitly find a transformation that takes one
to the other.

One of the problems in classical security is that information can
be copied: passwords can be stolen, songs can be pirated, and when
you email an attachment, you still have the original. One implication
is that E-commerce requires communicating with a server (e.g. the
credit card company or PayPal) whenever one makes a transaction. One
could hope that the no-cloning theorem would help circumvent this
and enable a physical quantum state to function like money. Such money
could be used in transactions both in person and on a future {}``Quantum
Internet'', not requiring contact with a central authority. As we
will see, quantum money is harder to forge than the paper bills in
our wallets. In order for a quantum state to function as money, we
require that:

\begin{enumerate}
\item The mint can produce it.
\item Anyone can verify it. That is, there is an efficient measurement that
anyone (with a quantum device) can perform on a quantum money state
that accepts genuine money with high probability and without significantly
damaging the money.
\item No one can forge it. That is, no one but the mint can efficiently
produce states that are accepted by the verifier with better than
exponentially small probability. In particular, it should not be possible
to copy a bill.
\end{enumerate}

A quantum money scheme has two components: pieces of quantum money
and an algorithm $M$ that verifies quantum money. A piece of quantum
money consists of a classical serial number $p$ (authenticated as
coming from the mint) along with an associated quantum state $|\$_{p}\rangle$
on $n$ qubits. The verification algorithm $M$ takes as input a quantum
state $|\phi\rangle$ and a serial number $q$ and then decides whether
or not the pair $\left(q,|\phi\rangle\right)$ is a piece of quantum
money. If the outcome is {}``good money'' then the verifier also
returns the state $|\phi\rangle$ undamaged so it can be used again.
We now formalize each of the above requirements:
\begin{enumerate}
\item \label{enu:There-is-an}There is a polynomial-time algorithm that
produces both a quantum money state $|\$_{p}\rangle$ and an associated
serial number $p$.
\item Running the verification algorithm $M$ with inputs $p$ and $|\$_{p}\rangle$
returns {}``good money'' and does not damage $|\$_{p}\rangle$.
Furthermore, anyone with access to a quantum computer (for example
a merchant) can run the verification algorithm $M$.
\item Given one piece of quantum money $\left(p,|\$_{p}\rangle\right)$,
it is hard to generate a quantum state $|\psi\rangle$ on $2n$ qubits
such that each part of $|\psi\rangle$ (along with the original serial
number $p$) passes the verification algorithm.
\end{enumerate}

What stops anyone other than the mint, using the same algorithm as
the mint, from producing counterfeit money states? This is why the
serial number needs to be authenticated. When the mint does a production
run it produces a set of pairs $\left(p,|\$_{p}\rangle\right)$. In
our quantum money scheme, the mint does not in advance choose the
serial numbers, but rather they are produced by a random process.
A rogue mint running the same algorithm as the mint can produce a
new set of money pairs, but (with high probability) none of the serial
numbers will match those that the mint originally produced. A simple
way to ensure security is for the mint to publish the list of valid
serial numbers and for the merchant to check against this list to
see if the serial number from the tendered money is authentic.

An alternative to publishing a list of serial numbers is for the money
state to come with a digital signature of the serial number. There
are known classical digital signature protocols that allow anyone
who knows the mint's {}``public key'' to verify that a serial number
is one that the mint did indeed produce. Still, the authenticated
description of the serial number can easily be copied and this is
where the quantum security comes into play: knowledge of the serial
number $p$ does not allow one to copy the associated quantum state
$|\$_{p}\rangle$. In the remainder of this paper, we assume that
anyone verifying quantum money will also check the authentication
of the serial number.

We can imagine two different ways to use quantum money for commerce.
If we had the technology to print a quantum state onto a piece of
paper, then we could use quantum money protocols to enhance the security
of paper money against forgery. Alternatively, people could use quantum
money states on their personal quantum computers to conduct business
either in person or over a quantum Internet. If small portable quantum
computers were available (imagine quantum smart phones or quantum
debit cards), then it would be easy to buy things with quantum money
instead of paper money.

For these uses, the {}``quantum money'' seen by an end-user would
either be a file on a quantum computer or a physical piece of money
with a quantum state somehow attached.

\subsection{Prior work}

The idea of quantum money was introduced by Wiesner in 1969, where
he proposed using the no cloning theorem to generate bills which can
not be copied \cite{BBBW83,Wie83}. However, in Wiesner's original
definition the only entity that could verify the quantum money state
was the mint that produced it. 

There is recent interest in designing quantum money which can be verified
by anyone who has a quantum computer (also called {}``public-key
quantum money''). Such money cannot be information-theoretically
secure; instead, it must rely on computational assumptions. Aaronson
showed that public-key quantum money exists relative to a quantum
oracle and proposed a concrete scheme without an oracle \cite{aaronson-quantum-money}.
Aaronson's concrete scheme was broken in \cite{LAF10}.

A different approach to quantum money was taken by Mosca and Stebila
\cite{mosca-2009,MS06,MS07,Ste09}, which is somewhat based on ideas
of Tokunaga, Okamoto, and Imoto \cite{TOI03}. This line of research
adds the additional requirement that all quantum money states be identical
and all share the same verification procedure. This additional requirement
is an abstraction of coins (which are identical) as opposed to bills
(which can have different serial numbers). In their works, Mosca and
Stebila give a way to generate quantum coins using a specialized random
quantum oracle. They also show how to generate quantum coins using
blind quantum computation. This requires the mint to be involved in
every transaction.

The main open question is how to design any secure quantum money (whether
of the quantum coin variety or otherwise) that does not require an
oracle and in which the mint's participation is not needed to use
the money.

A natural approach would be for the mint to choose some classical
secret and, using the secret, generate the money state and the serial
number. It should be hard to copy the quantum money state without
knowing the secret. One such proposal for a quantum money scheme is
based on product states. The mint first chooses a random product state
$|\$\rangle$ on $n$ qubits and then constructs a set of local projectors
$|\psi_{i}\rangle\langle\psi_{i}|$ for $i\in\{1,...m\}$ such that
$|\$\rangle$ is a zero eigenvector of each projector (and $m$ is
large enough that $|\$\rangle$ is the only state which is in the
common zero eigenspace). The serial number associated with the state
$|\$\rangle$ is a bit string which describes each of the projectors.
To verify a state $|\phi\rangle,$ one measures each projector $|\psi_{i}\rangle\langle\psi_{i}|$
and accepts the state if and only if the outcomes are all $0$. Unfortunately
this money scheme is insecure. It was shown in \cite{state-restoration}
that given a piece of product state quantum money $|\$\rangle$ and
the associated serial number it is possible to learn the classical
secret (the description of the product state). A generalized version
of this attack was also given in \cite{state-restoration}.

Due to the possibility of this type of attack, it was proposed in
\cite{LAF10} to use quantum money which is not based on a classical
secret, but rather on the hardness of generating a known superposition.
However, \cite{LAF10} did not present a proposal for a quantum money
scheme, only a blueprint.

\subsection{A blueprint for quantum money}

We present a scheme for quantum money which is not based on quantum
oracles and does not require communicating with the mint for verification.
The overall structure of the scheme is based on ideas from \cite{LAF10}.

The mint begins by generating a uniform superposition

\[
|\text{initial\ensuremath{\rangle}=}\frac{1}{\sqrt{|\mathcal{B}|}}\sum_{e\in\mathcal{B}}|e\rangle|0\rangle\]
 where $\mathcal{B}$ is a big set. It then applies some function
$f$ to $e$, obtaining the state 

\[
\frac{1}{\sqrt{|\mathcal{B}|}}\sum_{e\in\mathcal{B}}|e\rangle|f(e)\rangle.\]
 Finally, the mint measures the value of $f$. If the result of the
measurement is $v$, then the residual state is 

\[
|\$_{v}\rangle|v\rangle=\frac{1}{\sqrt{N_{v}}}\sum_{\substack{e\in\mathcal{B}\\
f(e)=v}
}|e\rangle|v\rangle.\]
The quantum money is the classical serial number $v$ and the quantum
state $|\$_{v}\rangle$.

To verify the money state, the verifier first measures the value of
$f$ on the state she is given to confirm that it has the value $v$.
Then she verifies that the state is in a uniform superposition of
states with the same value $v$ for $f$. It is not obvious in general
how to verify this uniform superposition and in our scheme we will
see that this is one of the main challenges.

For such a scheme to work, the following properties are required:
\begin{itemize}
\item Measuring $f$ twice on two unentangled copies of the initial state
should give different values with probability exponentially close
to 1. Otherwise, the forger can just repeat the procedure performed
by the mint, and obtain a money state with non negligible probability.
\item For {}``most'' values $v$, there should be exponentially many initial
states $e$ which have $f(e)=v$. Moreover, given one state $|e\rangle$
with $f(e)=v$ it should be hard to generate the uniform superposition
over all such states.
\item It should be possible for the verifier to verify that he has a uniform
superposition over states with the same value of $f$.
\end{itemize}

\subsection{Quantum money from knots}

To implement this kind of quantum money, we use ideas from knot theory.
The initial state generated by the mint is a uniform superposition
over planar grid diagrams, which are compact representations of oriented
links (an oriented knot is a knot with a preferred direction around
it, and an oriented link is a collection of possibly intertwined oriented
knots). The function $f$ which the mint measures is the Alexander
polynomial of the oriented link represented by a given planar grid
diagram. Finally, the verification procedure is based on Reidemeister
moves, which do not change the value of the polynomial. (Reidemeister
moves turn a knot into another equivalent knot, and the Alexander
polynomials for equivalent knots are the same.)

\section{Knots, links, and grid diagrams}

In this section we briefly review the standard concepts of knots and
links and how they are represented using diagrams. The same knot (or
link) can be represented as a diagram in many different ways; the
Reidemeister moves are a set of local changes to a diagram that do
not change the underlying knot or link. We will review how to compute
the Alexander polynomial for a given oriented link diagram in polynomial
time. The Alexander polynomial is a link invariant in the sense that
if we compute its value on diagrams which represent the same oriented
link we get the same polynomial. Finally, we review planar grid diagrams
and grid moves, which implement Reidemeister moves on grid diagrams.

\subsection{Knots and links\label{sub:Knots-and-Links}}

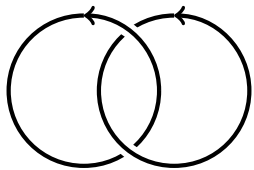
\begin{figure}[H]
\begin{centering}
\begin{tikzpicture}[ultra thick]

\def\outside{ (-1.2,-1.2) rectangle (2.4,1.2) }
\long\def\myclip#1#2{\begin{scope}[even odd rule] \clip \outside #1; #2\end{scope}}

\myclip{(1.2,-1.1) arc (270:180:1.1cm) -- (0.3,0) arc (180:270:0.9cm) -- cycle} {
 \draw[->] (0,1) arc (90:360:1cm) (1,0) arc (0:90:1cm);
}
\myclip{(0,1.1) arc (90:0:1.1cm) -- (0.9,0) arc (0:90:0.9cm) -- cycle} {
 \draw[->] (1.2,1) arc (90:360:1cm) arc (0:90:1cm);
}

\end{tikzpicture}
\par\end{centering}

\caption{An oriented link diagram.}
\label{Flo:linkdiagram}
\end{figure}

We can think of a knot as a loop of string in 3 dimensions, that is,
a map from $S^{1}$ into $\mathbb{R}^{3}$. Since it is hard to draw
knots in 3 dimensions, usually a knot is represented by a projection
of the 3 dimensional object into 2 dimensions where, at each crossing,
it is indicated which strand passes above and which below. This is
called a knot diagram. In this paper we will be interested in links,
which correspond to one or more loops of string (called the components
of the link). An oriented link is a link which has a direction associated
with each component. An example of an oriented link diagram is shown
in figure \ref{Flo:linkdiagram}.

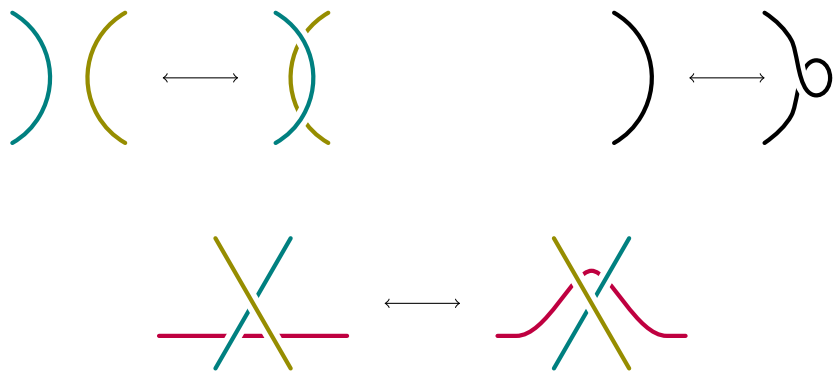
\begin{figure}[H]
\begin{centering}
\begin{tikzpicture}
 \draw[ultra thick,line cap=round,color=teal] (0,0.86603) arc (60:-60:1cm);
 \draw[ultra thick,line cap=round,color=olive] (1.5,0.86603) arc (120:240:1cm);
 \draw[<->] (2,0) -- (3,0);
 \begin{scope}[even odd rule]
  \clip (1,-1) rectangle (5,1) (3,0) circle (1.1cm) circle (0.9cm);
  \draw[ultra thick,line cap=round,color=olive] (4.2,0.86603) arc (120:240:1cm);
 \end{scope}
 \draw[ultra thick,line cap=round,color=teal] (3.5,0.86603) arc (60:-60:1cm);


\begin{scope}[shift={(8,0)}]
  \draw[ultra thick,line cap=round] (0,0.86603) arc (60:-60:1cm);

  \draw[<->] (1,0) -- (2,0);

\begin{scope}[shift={(2,0)}]
  \draw[ultra thick,line cap=round]
  (0.0000,0.8660)  .. controls (0.1527,0.7691) and (0.2824,0.6374) ..
  (0.3573,0.4993)  .. controls (0.4321,0.3616) and (0.4398,-0.0042) ..
  (0.5584,-0.1603) .. controls (0.6769,-0.3164) and (0.8713,-0.2029) ..
  (0.8713,0.0000);

\begin{scope}[even odd rule]
\clip (-1,-1) rectangle (1,1)
  (0.4688,-0.2188) .. controls (0.3874,-0.1116) and (0.3703,0.0256) ..
  (0.3438,0.1562) .. controls (0.3172,0.2869) and (0.3016,0.4000) ..
  (0.2812,0.4375) --
  (0.4375,0.5625) .. controls (0.4920,0.4623) and (0.5057,0.3132) ..
  (0.5312,0.1875) .. controls (0.5568,0.0618) and (0.5879,-0.0448) ..
  (0.6250,-0.0938) -- (0.4688,-0.2188) -- cycle;
\draw[ultra thick,line cap=round]
  (0.8713,0.0000)  .. controls (0.8713,0.2029) and (0.6769,0.3164) ..
  (0.5584,0.1603)  .. controls (0.4398,0.0042) and (0.4321,-0.3616) ..
  (0.3573,-0.4993) .. controls (0.2821,-0.6374) and (0.1527,-0.7601) ..
  (0.0000,-0.8660);
\end{scope}
\end{scope}
\end{scope}


\begin{scope}[shift={(2.7,-3)}]

\begin{scope}[ultra thick,line cap=round]

\def\outside{ (-1,-1) rectangle (3.0,1) }
\long\def\myclip#1#2{\begin{scope}[even odd rule] \clip \outside #1; #2\end{scope}}
\def\urclip{(0,0.86603-0.2) -- (1,-0.86603-0.2) -- (1,-0.86603+0.2) -- (0,0.86603+0.2) -- cycle}
\def\ulclip{(0,-0.86603-0.2) -- (1,0.86603-0.2) -- (1,0.86603+0.2) -- (0,-0.86603+0.2) -- cycle}

\myclip{\urclip}{\draw[color=teal] (0,-0.86603) -- (1,0.86603);}
\draw[color=olive] (0,0.86603) -- (1,-0.86603);
\myclip{\urclip \ulclip}{\draw[color=purple] (-0.75,-0.4330) -- (1.75,-0.4330);}

\end{scope}

\draw[<->] (2.25,0) -- (3.25,0);

\begin{scope}[shift={(4.5,0)},ultra thick,line cap=round]

\def\outside{ (-1,-1) rectangle (3.0,1) }
\long\def\myclip#1#2{\begin{scope}[even odd rule] \clip \outside #1; #2\end{scope}}
\def\urclip{(0,0.86603-0.2) -- (1,-0.86603-0.2) -- (1,-0.86603+0.2) -- (0,0.86603+0.2) -- cycle}
\def\ulclip{(0,-0.86603-0.2) -- (1,0.86603-0.2) -- (1,0.86603+0.2) -- (0,-0.86603+0.2) -- cycle}

\myclip{\urclip}{\draw[color=teal] (0,-0.86603) -- (1,0.86603);}
\draw[color=olive] (0,0.86603) -- (1,-0.86603);
\myclip{\urclip \ulclip}{
  \draw[color=purple] (-0.75,-0.4330)
    -- (-.5,-0.4330) .. controls (-0.1,-0.4330)
    and (0.25,0.4330) .. (0.5,0.4330) .. controls (0.75,0.4330)
    and (1.1,-0.4330) .. (1.5,-0.4330)
    -- (1.75,-0.4330);
}

\end{scope}
\end{scope}
\end{tikzpicture}
\par\end{centering}

\caption{The three Reidemeister moves for unoriented link diagrams. }
\label{Flo:Reidemeister}
\end{figure}

 Two links (or knots) are equivalent if one can be smoothly morphed
into the other without cutting the string. If unoriented links $K_{1}$
and $K_{2}$ are equivalent and they are represented by diagrams $D_{1}$
and $D_{2}$ respectively, then diagram $D_{1}$ can be transformed
into $D_{2}$ (and vice versa) using the Reidemeister moves pictured
in figure \ref{Flo:Reidemeister}. (For oriented links the Reidemeister
moves can be drawn with the same pictures as in figure \ref{Flo:Reidemeister}
but with orientations along the edges which have to be consistent
before and after applying the move). Looking at these moves, one sees
that if two diagrams can be brought into the same form by using the
moves then the diagrams represent equivalent links. The converse of
this statement is a theorem.

\subsection{The Alexander polynomial of an oriented link\label{sub:The-Alexander-polynomial}}

The Alexander polynomial is a polynomial $\Delta(x)$ which can be
computed from a given oriented link diagram and which is invariant
under the Reidemeister moves. In this section, following Alexander
\cite{1928}, we describe how to compute this polynomial. It will
be clear from our discussion that the computation of $\Delta(x)$
can be done in polynomial time in the number of crossings of the diagram.

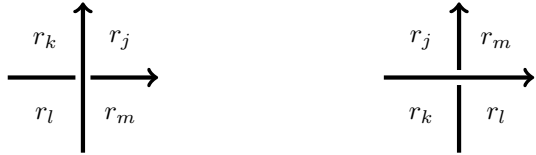
\begin{figure}[H]
\begin{centering}
\begin{tikzpicture}[ultra thick]

\begin{scope}
 \draw (-1,0) -- (-0.1,0);
 \draw[->] (0.1,0) -- (1,0);
 \draw[->] (0,-1) -- (0,1);
 \node at (-0.5,0.5) {$r_k$};
 \node at (0.5,0.5) {$r_j$};
 \node at (-0.5,-0.5) {$r_l$};
 \node at (0.5,-0.5) {$r_m$};
\end{scope}

\begin{scope}[shift={(5,0)}]
 \draw (0,-1) -- (0,-0.1);
 \draw[->] (0,0.1) -- (0,1);
 \draw[->] (-1,0) -- (1,0);
 \node at (-0.5,0.5) {$r_j$};
 \node at (0.5,0.5) {$r_m$};
 \node at (-0.5,-0.5) {$r_k$};
 \node at (0.5,-0.5) {$r_l$};
\end{scope}
\end{tikzpicture}
\par\end{centering}

\caption{An equation $xr_{j}-xr_{k}+r_{l}-r_{m}=0$ is associated with each
crossing, where the adjacent regions $\{r_{j},r_{k},r_{l},r_{m}\}$
are labeled as shown. Note that the labeling of the adjacent regions
depends on the which line crosses on top. }
\label{Flo:labelingregions}
\end{figure}

Suppose we are given a diagram of an oriented link $L$. If the diagram
is disconnected, apply the first Reidemeister move in Figure~\ref{Flo:Reidemeister}
to connect the diagram. Let us write $a$ for the number of crosses
in the diagram. The curve of the diagram then divides the two dimensional
plane into $a+2$ regions including one infinite region (this follows
from Euler's formula). The following recipe can be used to calculate
$\Delta(x)$:
\begin{enumerate}
\item For each region $i\in\{1,...,a+2\}$, associate a variable $r_{i}$.
\item For each of the $a$ crossings, write down an equation\[
xr_{j}-xr_{k}+r_{l}-r_{m}=0,\]
 where $\{r_{j},r_{k},r_{l},r_{m}\}$ are the variables associated
with the regions adjacent to the crossing, in the order pictured in
figure \ref{Flo:labelingregions}.
\item Write the set of equations as a matrix equation \[
\mathcal{M}\left(\begin{array}{c}
r_{1}\\
r_{2}\\
\vdots\\
r_{m+2}\end{array}\right)=0.\]
 This defines the matrix $\mathcal{M}$ which has $a$ rows and $a+2$
columns. The entries of $\mathcal{M}$ are elements of the set $\{\pm1,\pm x,1+x,1-x,-1+x,-1-x\}.$
\item Delete two columns of $\mathcal{M}$ which correspond to adjacent
regions in the link diagram. This gives an $a\times a$ matrix $\mathcal{M}_{0}$.
\item Take the determinant of $\mathcal{M}_{0}$. This is a polynomial in
$x$. Divide this polynomial by the factor $\pm x^{q}$ chosen to
make the lowest degree term a positive constant. The resulting polynomial
is the Alexander polynomial $\Delta\left(x\right)$ for the given
link.
\end{enumerate}
When we use the Alexander polynomial in this paper, we are referring
to the list of coefficients, not the value of the polynomial evaluated
at some $x$.

\subsection{Grid diagrams\label{sub:Grid-diagrams}}

It is convenient to represent knots as \emph{planar grid diagrams}.
A planar grid diagram is a $d\times d$ grid on which we draw $d$
X's and $d$ O's. There must be exactly one X and one O in each row
and in each column, and there may never be an X and an O in the same
cell. We draw a horizontal line in each row between the O and the
X in that row and a vertical line in each column between the X and
the O in that column. Where horizontal lines and vertical lines intersect,
the vertical line always crosses above the horizontal line.

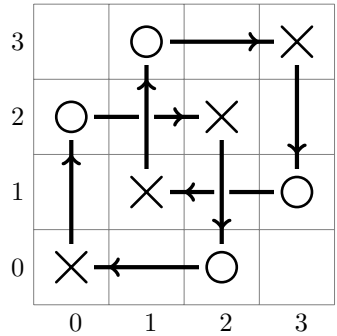
\begin{figure}[H]
\begin{centering}
\begin{tikzpicture}
  \griddims(4,4)
  \draw[help lines] (0, 0) grid (4, 4);
  \xlabel(0,$0$) \xlabel(1,$1$) \xlabel(2,$2$) \xlabel(3,$3$)
  \ylabel(0,$0$) \ylabel(1,$1$) \ylabel(2,$2$) \ylabel(3,$3$)
  \startmarks
    \vseg(0,0,2) \vseg(1,1,3) \vseg(2,0,2) \vseg(3,1,3)
    \uarrow(0,1) \uarrow(1,2) \darrow(2,1) \darrow(3,2)
    \rarrow(1,2) \rarrow(2,3) \larrow(1,0) \larrow(2,1)
    \hseg(0,2,0) \hseg(1,3,3)
    \clipseg{\crossgap(1,2)}{\hseg(0,2,2)}
    \clipseg{\crossgap(2,1)}{\hseg(1,3,1)}
    \redmark(0,0) \redmark(1,1) \redmark(2,2) \redmark(3,3)
    \bluemark(0,2) \bluemark(2,0) \bluemark(1,3) \bluemark(3,1)
  \endmarks
\end{tikzpicture}

\par\end{centering}

\caption{A simple planar grid diagram}
\label{fig:simple-grid-diag}
\end{figure}

Knots (or links) in a grid diagram carry an implicit orientation:
each vertical edge goes from an X to an O, and each horizontal edge
goes from an O to an X. Figure~\ref{fig:simple-grid-diag} shows
an example of a $d=4$ planar grid diagram. 

A planar grid diagram $G$ can be specified by two disjoint permutations
$\pi_{X},\pi_{O}\in S_{d}$, in which case the X's have coordinates
$\{(i,\pi_{X}(i))\}$ and the O's have coordinates $\{(i,\pi_{O}(i))\}$
for $i\in\{1,...,d\}$. Two permutations are said to be disjoint if
for all $i$, $\pi_{X}(i)\neq\pi_{O}(i)$. Any two disjoint permutations
$\pi_{X},\pi_{O}\in S_{d}$ thus define a planar grid diagram $G=(\pi_{X},\pi_{O})$.
Every link can be represented by many different grid diagrams.

\begin{figure}[H]
\begin{centering}
\begin{tikzpicture}

 \griddims(4,6)

 \begin{scope}[even odd rule]
  \clipgrid(0,0, 4,6)
  \iclipgrid(1,1, 3,5, 0,0, 4,6)
  \draw[help lines,dashed] (0,0) grid (4,6);
 \end{scope}
 \draw (1,1) grid (3,5);

 \xlabel(1,$e$) \xlabel(2,$e+1$)

 \startmarks
   \hsegfl(0,1,1) \hsegfl(0,1,4)
   \vseg(1,1,4) \darrow(1,2)
   \vseg(2,2,3) \uarrow(2,2)
   \clipseg{\crossgap(1,3)}{\hsegfl(0,2,3)}
   \clipseg{\crossgap(1,2)}{\hsegfl(0,2,2)}
   \redmark(1,4) \bluemark(1,1) \redmark(2,2) \bluemark(2,3)
 \endmarks

\draw[<->] (4.5,3) -- (5.5,3);

\begin{scope}[shift={(6,0)}]
 \griddims(4,6)

 \begin{scope}[even odd rule]
  \clipgrid(0,0, 4,6)
  \iclipgrid(1,1, 3,5, 0,0, 4,6)
  \draw[help lines,dashed] (0,0) grid (4,6);
 \end{scope}
 \draw (1,1) grid (3,5);

 \xlabel(1,$e$) \xlabel(2,$e+1$)

 \startmarks
   \hsegfl(0,2,1) \hsegfl(0,2,4)
   \vseg(2,1,4) \darrow(2,2)
   \vseg(1,2,3) \uarrow(1,2)
   \hsegfl(0,1,3)
   \hsegfl(0,1,2)
   \redmark(2,4) \bluemark(2,1) \redmark(1,2) \bluemark(1,3)
 \endmarks

\end{scope}


\begin{scope}[shift={(0,-6.75)}]

 \griddims(4,6)

 \begin{scope}[even odd rule]
  \clipgrid(0,0, 4,6)
  \iclipgrid(1,1, 3,5, 0,0, 4,6)
  \draw[help lines,dashed] (0,0) grid (4,6);
 \end{scope}
 \draw (1,1) grid (3,5);

 \xlabel(1,$e$) \xlabel(2,$e+1$)

 \startmarks
   \hsegfl(0,1,1)
   \vseg(1,1,3) \darrow(1,2)
   \vseg(2,2,4) \uarrow(2,3)
   \hsegfr(2,3,2)
   \hsegfl(0,2,4)
   \clipseg{\crossgap(2,3)}{\hsegfr(1,3,3)}
   \redmark(1,3) \bluemark(1,1) \redmark(2,2) \bluemark(2,4)
 \endmarks

\draw[<->] (4.5,3) -- (5.5,3);

\begin{scope}[shift={(6,0)}]

\begin{scope}[even odd rule]
  \clipgrid(0,0, 4,6)
  \iclipgrid(1,1, 3,5, 0,0, 4,6)
  \draw[help lines,dashed] (0,0) grid (4,6);
 \end{scope}
 \draw (1,1) grid (3,5);

 \xlabel(1,$e$) \xlabel(2,$e+1$)

 \startmarks
   \hsegfl(0,2,1)
   \vseg(2,1,3) \darrow(2,2)
   \vseg(1,2,4) \uarrow(1,3)
   \clipseg{\crossgap(2,2)}{\hsegfr(1,3,2)}
   \hsegfl(0,1,4)
   \hsegfr(2,3,3)
   \redmark(2,3) \bluemark(2,1) \redmark(1,2) \bluemark(1,4)
 \endmarks

\end{scope}

\draw[red,line cap=round,thick] (-0.5,0) -- (10.5,6) (10.5,0) -- (-0.5,6);

\end{scope}


\begin{scope}[shift={(0,-13.5)}]

 \griddims(4,6)

 \begin{scope}[even odd rule]
  \clipgrid(0,0, 4,6)
  \iclipgrid(1,1, 3,5, 0,0, 4,6)
  \draw[help lines,dashed] (0,0) grid (4,6);
 \end{scope}
 \draw (1,1) grid (3,5);

 \xlabel(1,$e$) \xlabel(2,$e+1$)

 \startmarks
   \hsegfl(0,1,1)
   \vseg(1,1,2) \darrow(1,2)
   \vseg(2,3,4) \uarrow(2,3)
   \hsegfr(1,3,2)
   \hsegfl(0,2,4)
   \hsegfr(2,3,3)
   \redmark(1,2) \bluemark(1,1) \redmark(2,3) \bluemark(2,4)
 \endmarks

\draw[<->] (4.5,3) -- (5.5,3);

\begin{scope}[shift={(6,0)}]

\griddims(4,6)

 \begin{scope}[even odd rule]
  \clipgrid(0,0, 4,6)
  \iclipgrid(1,1, 3,5, 0,0, 4,6)
  \draw[help lines,dashed] (0,0) grid (4,6);
 \end{scope}
 \draw (1,1) grid (3,5);

 \xlabel(1,$e$) \xlabel(2,$e+1$)

 \startmarks
   \hsegfl(0,2,1)
   \vseg(2,1,2) \darrow(2,2)
   \vseg(1,3,4) \uarrow(1,3)
   \hsegfr(2,3,2)
   \hsegfl(0,1,4)
   \hsegfr(1,3,3)
   \redmark(2,2) \bluemark(2,1) \redmark(1,3) \bluemark(1,4)
 \endmarks

\end{scope}

\end{scope}

\end{tikzpicture}

\par\end{centering}

\caption{Transposition of adjacent columns $e$ and $e+1$. The legality of
these moves depends only on the positions of markers in the columns
being transposed. The positions of markers in other columns are irrelevant.\label{fig:grid-transposition}
The middle move is not allowed.}

\end{figure}
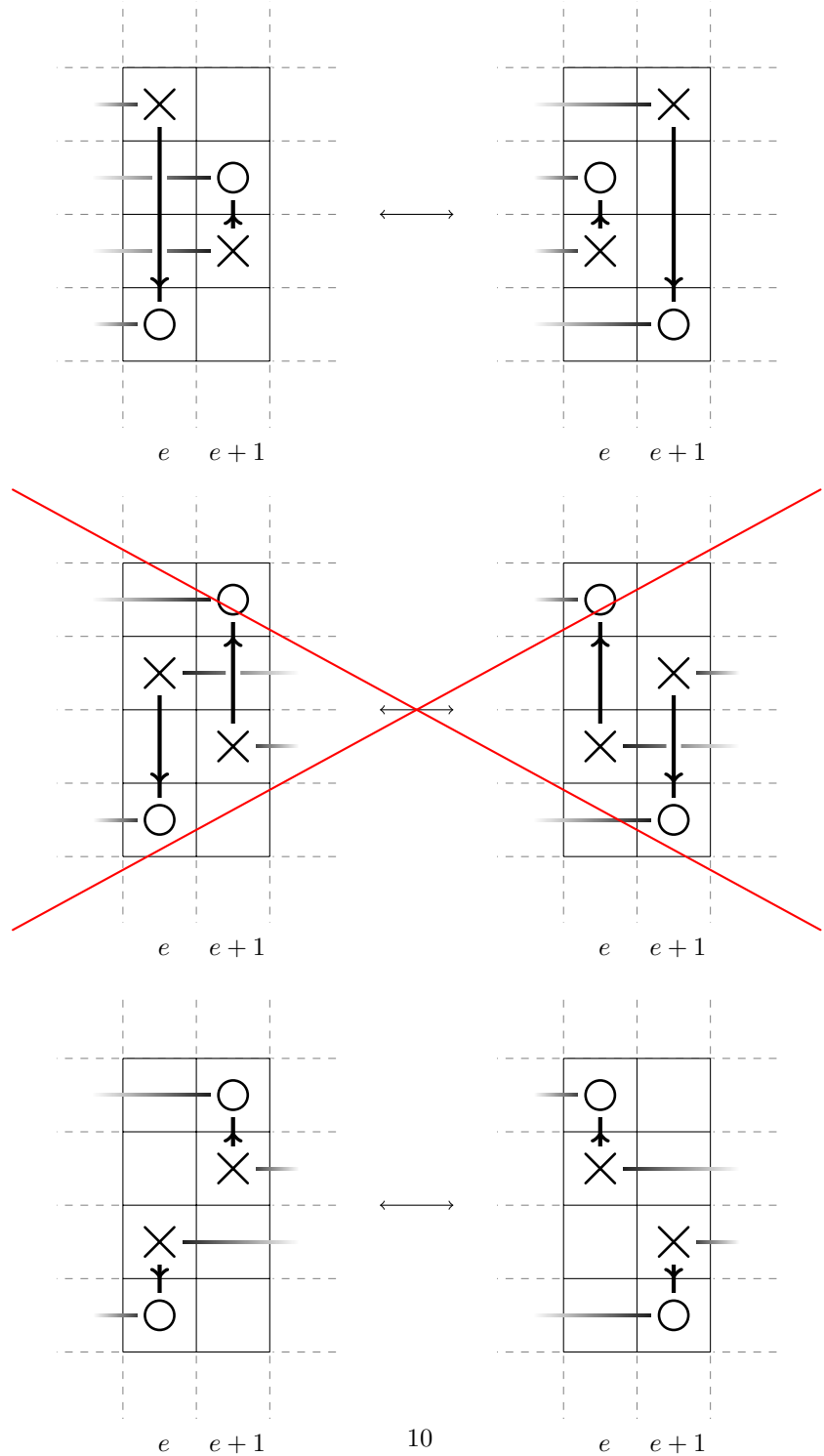

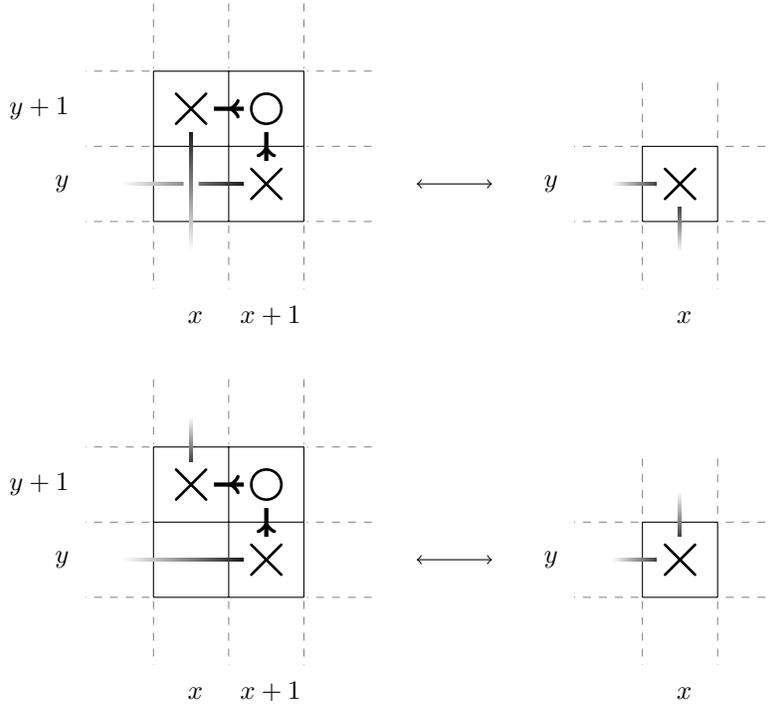
\begin{figure}[H]
\begin{centering}
\begin{tikzpicture}

 \griddims(4,4)

 \begin{scope}[even odd rule]
  \clipgrid(0,0, 4,4)
  \iclipgrid(1,1, 3,3)
  \draw[help lines,dashed] (0,0) grid (4,4);
 \end{scope}
 \draw (1,1) grid (3,3);

 \xlabel(1,$x$) \xlabel(2,$x+1$)
 \ylabel(1,$y$) \ylabel(2,$y+1$)

 \startmarks
   \vsegfd(1,0,2)
   \hseg(1,2,2) \larrow(2,2)
   \vseg(2,1,2) \uarrow(2,1)
   \clipseg{\crossgap(1,1)}{\hsegfl(0,2,1)}
   \redmark(1,2) \redmark(2,1) \bluemark(2,2)
 \endmarks

\draw[<->] (4.5,1.5) -- (5.5,1.5);

\begin{scope}[shift={(6.5,0)}]
 \griddims(3,3)

 \begin{scope}[even odd rule]
  \clipgrid(0,0, 3,3)
  \iclipgrid(1,1, 2,2)
  \draw[help lines,dashed] (0,0) grid (3,3);
 \end{scope}
 \draw (1,1) grid (2,2);

 \xlabel(1,$x$)
 \ylabel(1,$y$)

 \startmarks
   \vsegfd(1,0,1)
   \hsegfl(0,1,1)
   \redmark(1,1)
 \endmarks

\end{scope}


\begin{scope}[shift={(0,-5)}]

 \griddims(4,4)

 \begin{scope}[even odd rule]
  \clipgrid(0,0, 4,4)
  \iclipgrid(1,1, 3,3)
  \draw[help lines,dashed] (0,0) grid (4,4);
 \end{scope}
 \draw (1,1) grid (3,3);

 \xlabel(1,$x$) \xlabel(2,$x+1$)
 \ylabel(1,$y$) \ylabel(2,$y+1$)

 \startmarks
   \vsegfu(1,2,3)
   \hseg(1,2,2) \larrow(2,2)
   \vseg(2,1,2) \uarrow(2,1)
   \hsegfl(0,2,1)
   \redmark(1,2) \redmark(2,1) \bluemark(2,2)
 \endmarks

\draw[<->] (4.5,1.5) -- (5.5,1.5);

\begin{scope}[shift={(6.5,0)}]
 \griddims(3,3)

 \begin{scope}[even odd rule]
  \clipgrid(0,0, 3,3)
  \iclipgrid(1,1, 2,2)
  \draw[help lines,dashed] (0,0) grid (3,3);
 \end{scope}
 \draw (1,1) grid (2,2);

 \xlabel(1,$x$)
 \ylabel(1,$y$)

 \startmarks
   \vsegfu(1,1,2)
   \hsegfl(0,1,1)
   \redmark(1,1)
 \endmarks

\end{scope}
\end{scope}

\end{tikzpicture}

\par\end{centering}

\caption{Stabilization (right to left) and destabilization (left to right).
This move is legal when rotated as well as when the markers are switched
between X and O. The position of other markers does not matter.\label{fig:grid-stabilization}}

\end{figure}

We can define grid moves, which are three types of moves (that is,
transformations) on planar grid diagrams which, like the Reidemeister
moves for link diagrams, are sufficient to generate all planar grid
diagrams of the same oriented link.
\begin{itemize}
\item The first type of move is a cyclic permutation of either the rows
or the columns. Figure~\ref{fig:grid-cyclic-perm} shows an example
of this type of move on columns. We can think of these moves as grabbing
both markers in the rightmost column, pulling them behind of the page,
and putting them back down on the left. These moves are always legal.
There are equivalent moves on rows.
\item The second type of move is transposition of two adjacent rows or columns.
This can be done only when no strand would be broken. In Figure~\ref{fig:grid-transposition}
we show examples of when this move is allowed. The legality of this
move depends only on the position of the markers in the rows or columns
being transposed.%
\footnote{For the case of columns $i$ and $i+1$, the precise condition is
that either the two intervals $\left[\min\left(x_{i},o_{i}\right),\max\left(x_{i},o_{i}\right)\right]$
and $\left[\min\left(x_{i+1},o_{i+1}\right),\max\left(x_{i+1},o_{i+1}\right)\right]$
do not overlap or one of the intervals contains the other.%
}
\item The third type of move adds one row and one column (this is called
stabilization) or deletes one row and one column (this is called destabilization),
as shown in Figure~\ref{fig:grid-stabilization}. Destabilization
selects three markers forming an {}``L'' shape with sides of length
1 and collapses them into one. That is, it deletes one row and one
column such that all three markers are removed. The inverse move selects
any marker, adds a row and a column adjacent to that marker, and replaces
that marker with three new markers. Any X or O can always be legally
stabilized and any three markers forming an {}``L'' shape with sides
of length 1 can be destabilized unless they form a box (i.e. a $2\times2$
square with a marker in all four positions).
\end{itemize}
In the remainder of this paper we will represent links exclusively
with planar grid diagrams.

\begin{figure}[H]
\begin{centering}
\begin{tikzpicture}

 \griddims(5,6)

 \begin{scope}[even odd rule]
  \clipgrid(0,0, 5,6)
  \iclipgrid(0,1, 1,5)
  \iclipgrid(4,1, 5,5)
  \draw[help lines,dashed] (0, 0) grid (5, 6);
 \end{scope}
 \draw (0, 1) grid (1, 5);
 \draw (4, 1) grid (5, 5);

 \xlabel(0,$0$) \xlabel(4,$d-1$)

 \startmarks
   \hsegfr(0,3,1) \hsegfr(0,3,3) \hsegfl(1,4,2) \hsegfl(1,4,4)
   \vseg(4,2,4) \uarrow(4,3)
   \vseg(0,1,3) \uarrow(0,2)
   \redmark(0,1) \bluemark(0,3) \redmark(4,2) \bluemark(4,4)
 \endmarks

\draw[<->] (5.5,3) -- (6.5,3);

\begin{scope}[shift={(7,0)}]

\begin{scope}[even odd rule]
  \clipgrid(0,0, 5,6)
  \iclipgrid(0,1, 2,5)
  \draw[help lines,dashed] (0, 0) grid (5, 6);
 \end{scope}
 \draw (0, 1) grid (2, 5);

 \xlabel(0,$0$) \xlabel(1,$1$)

 \startmarks
   \hsegfr(1,4,1) \hsegfr(1,4,3) \hsegfr(2,0,4)
   \begin{scope}[even odd rule] \clip (0,0) rectangle (5,6) (1-0.1,2-0.1) rectangle (1+0.1,2+0.1); \hsegfr(2,0,2) \end{scope}
   \vseg(0,2,4) \uarrow(0,3)
   \vseg(1,1,3) \uarrow(1,2)
   \redmark(1,1) \bluemark(1,3) \redmark(0,2) \bluemark(0,4)
 \endmarks

\end{scope}

\end{tikzpicture}

\par\end{centering}

\caption{Cyclic permutation of columns. This move is always legal.\label{fig:grid-cyclic-perm}}

\end{figure}
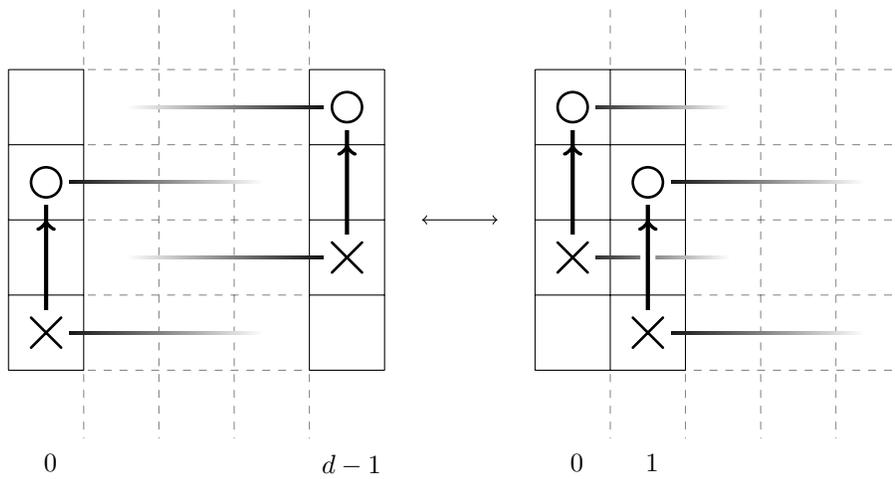

\section{Quantum money\label{sec:Quantum-money}}

In this section we describe our quantum money scheme in full detail.
In section \ref{sub:Minting-Quantum-Money} we describe how the mint
can make quantum money efficiently. In section \ref{sub:Verifying-quantum-money}
we describe the verification procedure which can be used by anyone
(that is, anyone who owns a quantum computer) to ensure that a given
quantum bill is legitimate. In section \ref{sub:Discussion} we discuss
the security of the verification procedure. In section \ref{sub:Why-not-Quantum}
we contrast our quantum money scheme based on knots with a similar
but insecure scheme based on graphs.

\subsection{Minting quantum money\label{sub:Minting-Quantum-Money}}

The basis vectors of our Hilbert space are grid diagrams $|G\rangle=|\pi_{X},\pi_{O}\rangle$,
where each grid diagram $G$ is encoded by the disjoint permutations
$\pi_{X},\pi_{O}$. The dimension $d(G)$ is the number of elements
in the permutations.%
\footnote{In practice, we will encode $d$, $\pi_{X}$ and $\pi_{O}$ as bit
strings. Any such encoding will also contain extraneous bit strings
that do not describe diagrams; the verification algorithm will ensure
that these do not occur.%
} The number of disjoint pairs of permutations on $d$ elements is
equal to $d!$ times the number of permutations that have no fixed
points. The latter quantity is called the number of derangements on
$d$ elements, and it is given by $\left[\frac{d!}{e}\right]$ where
the brackets indicate the nearest integer function.

To generate quantum money, the mint first chooses a security parameter
$\bar{D}$ and defines an unnormalized distribution \[
y\left(d\right)=\begin{cases}
\frac{1}{d!\left[\frac{d!}{e}\right]}\exp\frac{-\left(d-\bar{D}\right)^{2}}{2\bar{D}} & \text{ if }2\le d\le2\bar{D}\\
0 & \text{ otherwise}.\end{cases}\]
Define an integer valued function \[
q(d)=\lceil\frac{y(d)}{y_{min}}\rceil,\]
where $\lceil\cdot\rceil$ means round up and where $y_{min}$ is
the minimum value of $y(d)$ for $d\in{2,...,2\bar{D}}$ (we need
an integer valued distribution in what follows). The mint then uses
the algorithm in \cite{grover-2002} to prepare the state (up to normalization)
\[
\sum_{d=2}^{2\bar{D}}d!\sqrt{q(d)}|d\rangle.\]
Using a straightforward unitary transformation acting on this state
and an additional register, the mint produces \[
\sum_{d=2}^{2\bar{D}}d!\sqrt{q\left(d\right)}|d\rangle\bigg(\frac{1}{d!}\sum_{\pi_{X},\pi_{O}\in S_{d}}|\pi_{X},\pi_{O}\rangle\bigg),\]
and then measures whether or not the two permutations $\pi_{X}$ and
$\pi_{O}$ are disjoint. (They are disjoint with probability very
close to $\nicefrac{1}{e}$.) If the mint obtains the measurement
outcome {}``disjoint'', it uncomputes the dimension $d$ in the
first register to obtain the state $|\text{initial}\rangle$ on the
last two registers, where \begin{equation}
|\text{initial}\rangle=\frac{1}{\sqrt{N}}\sum_{\text{grid diagrams }G}\sqrt{q\left(d\left(G\right)\right)}|G\rangle\label{eq:initialstate}\end{equation}
and $N$ is a normalizing constant. If the measurement says {}``not
distinct'', the mint starts over.

The distribution $q\left(d\right)$ is chosen so that if one were
to measure $d(G)$ on $|\text{initial}\rangle$, then the distribution
of results would be extremely close to Gaussian with mean $\bar{D}$
restricted to integers in the interval $\left[2,2\bar{D}\right]$.
As $\bar{D}$ becomes large, the missing weight in the tails falls
rapidly to zero.

From the state $|\text{initial}\rangle$, the mint computes the Alexander
polynomial $A(G)$ into another register and then measures this register,
obtaining the polynomial $p$. The resulting state is $|\text{\$}_{p}\rangle$,
the weighted superposition of all grid diagrams (up to size $2\bar{D}$)
with Alexander polynomial $p$: \begin{equation}
|\text{\$}_{p}\rangle=\frac{1}{\sqrt{N}}\sum_{G:A\left(G\right)=p}\sqrt{q(d(G))}|G\rangle,\label{eq:moneystate}\end{equation}
where $N$ takes care of the normalization. The quantum money consists
of the state $|\$_{p}\rangle$, and the serial number is the polynomial
$p$, represented by its list of coefficients. If the polynomial $p$
is zero, the mint should reject that state and start over. Splittable
links have Alexander polynomial equal to zero, and we wish to avoid
this case.

\subsection{Verifying quantum money\label{sub:Verifying-quantum-money}}

Suppose you are a merchant. Someone hands you a quantum state $|\phi\rangle$
and a serial number which corresponds to a polynomial $p$ and claims
that this is good quantum money. To check that indeed this is the
case, you would use the following algorithm.

\begin{algorithm}[H]
$\hphantom{}$
\begin{enumerate}
\item [0.]\setcounter{enumi}{0}Verify that $|\phi\rangle$ is a superposition
of basis vectors that validly encode grid diagrams. If this is the
case then move on to step \ref{enu:Measure-the-Alexander}, otherwise
reject.
\item \label{enu:Measure-the-Alexander}Measure the Alexander polynomial
on the state $|\phi\rangle$. If this is measured to be $p$ then
continue on to step \ref{enu:Measure-the-projector}. Otherwise, reject.
\item \label{enu:Measure-the-projector}Measure the projector onto grid
diagrams with dimensions in the range $[\frac{\bar{D}}{2},\frac{3\bar{D}}{2}].$
If you obtain $+1$ then continue on to step \ref{enu:Apply-the-Markov}.
Otherwise, reject. For valid money, you will obtain the outcome $+1$
with high probability and cause little damage to the state. (We discuss
the rationale for this step in section~\ref{sub:Discussion}.)
\item \label{enu:Apply-the-Markov}Apply the Markov chain verification algorithm
described in section \ref{sub:A-Quantum-Verification}. If the state
passes this step, accept the state. Otherwise, reject.
\end{enumerate}
\caption{Verifying Quantum Money}

\end{algorithm}

If the money passes steps 0 and \ref{enu:Measure-the-Alexander} then
$|\phi\rangle$ is a superposition of grid diagrams with the correct
Alexander polynomial $p$. Now passing steps \ref{enu:Measure-the-projector}
and \ref{enu:Apply-the-Markov} will (approximately) confirm that
$|\phi\rangle$ is the \emph{correct} superposition of grid diagrams.
This procedure will accept genuine quantum money states with high
probability, but, as we discuss below, there will be other states
that also pass verification. We believe that these states are hard
to manufacture.

We now discuss the quantum verification scheme used in step \ref{enu:Apply-the-Markov}.
We begin by defining a classical Markov chain.

\subsubsection{A classical markov chain\label{sub:A-Markov-Chain}}

The Markov chain is chosen to have uniform limiting distribution over
pairs $(G,i)$ where $G$ is equivalent to (and reachable without
exceeding grid dimension $2\bar{D}$ from) the starting configuration,
and where $i\in\{1,...,q(d(G))\}.$ Therefore, in the limiting distribution
(starting from a grid diagram $\tilde{G}$) the probability of finding
a grid diagram $G$ (which is equivalent to $\tilde{G}$) is proportional
to $q(d(G)).$ We use the extra label $i$ in our implementation of
the verifier.

There are two types of update rules for the Markov chain. The first
type is an update that changes $i$ while leaving $G$ unchanged (the
new value of $i$ is chosen uniformly). The second type is an update
that changes $G$ to a new diagram $G^{\prime}$ while leaving $i$
alone (this type of update is only allowed if $i\leq q(d(G^{\prime}))$).
For the moves which leave $i$ unchanged, our Markov chain selects
a grid move at random and proposes to update the current grid diagram
by applying that move. The move $G\rightarrow G^{\prime}$ is accepted
if the value $i\leq q(d(G^{\prime})).$ 

Recall (from section \ref{sub:Grid-diagrams}) that there is a set
of grid moves on planar grid diagrams which can be used to transition
from one grid diagram to another. These moves only allow transitions
between grid diagrams which represent equivalent links, and some of
these moves change the dimension of the grid diagram. In general,
any two grid diagrams representing equivalent links can be connected
by a sequence of these moves (although finding this sequence is not
easy). However, this sequence of moves may also pass through grid
diagrams which have dimension greater than $2\bar{D}$. In this case,
the two equivalent diagrams will not mix, due to the cutoff we defined.

In the appendix we define this Markov chain in detail. This Markov
chain does not mix between nonequivalent grid diagrams. As such, it
has many stationary distributions. In the next section, we use this
Markov chain to produce a quantum verification procedure.

\subsubsection{A quantum verification procedure based on a classical Markov chain
\label{sub:A-Quantum-Verification}}

Let $B$ denote the Markov matrix. As shown in the appendix, $B$
is a doubly stochastic (the sum of the elements in each row and each
column is $1$) matrix which can be written as \begin{equation}
B=\frac{1}{\left|\mathcal{S}\right|}\sum_{s\in\mathcal{S}}P_{s},\label{eq:B_matrix}\end{equation}
where each $P_{s}$ is a permutation on the configuration space of
pairs $(G,i)$ such that $i\leq q(d(G))$, and $\mathcal{S}$ is a
set of moves that the Markov chain can make at each iteration. We
can view the same matrices as quantum operators acting on states in
a Hilbert space where basis vectors are of the form $|G\rangle|i\rangle$.
For clarity we will refer to the quantum operators as $\hat{B}$ and
$\hat{P}_{s}$.

Our quantum money states $|\$_{p}\rangle$ live in a Hilbert space
where basis vectors are grid diagrams $|G\rangle.$ To enable our
verification procedure, we define a related state $|\$_{p}^{\prime}\rangle$
on two registers, where the second register holds integers $i\in\{0,...,q_{max}\}$
(here $q_{max}$ is the maximum that $q(d)$ can reach). $|\$'_{p}\rangle$
is unitarily related to $|\$_{p}\rangle$, so verifying the former
is equivalent to verifying the latter. Define a unitary $U$ which
acts on basis vectors in the expanded Hilbert space as\[
U\left(|G\rangle|0\rangle\right)=|G\rangle\frac{1}{\sqrt{q(d(G))}}\sum_{i=1}^{q(d(G))}|i\rangle.\]
To obtain $|\$_{p}^{\prime}\rangle,$ take the quantum money state
$|\$_{p}\rangle$ and adjoin the ancilla register initialized in the
state $|0\rangle$. Then apply $U$ to both registers to produce the
state \begin{eqnarray*}
|\$_{p}^{\prime}\rangle & = & U\left(|\$_{p}\rangle|0\rangle\right)\\
 & = & \frac{1}{\sqrt{N}}\sum_{G:A(G)=p}\sum_{i=1}^{q(d(G))}|G\rangle|i\rangle.\end{eqnarray*}
 Note that whereas $|\$_{p}\rangle$ is a weighted superposition in
the original Hilbert space (see equation \ref{eq:moneystate}), $|\$_{p}^{\prime}\rangle$
is a uniform superposition in the expanded Hilbert space. However
both $|\$_{p}\rangle$ and $|\$_{p}^{\prime}\rangle$ give rise to
the same probability distribution over grid diagrams $G$. 

The subroutine in step \ref{enu:Apply-the-Markov} of the verification
procedure of Section \ref{sub:Verifying-quantum-money} starts with
the input state $|\phi\rangle$ and applies the following algorithm.%
\footnote{Using a construction from \cite{aharonov-2003}, it is possible to
associate a Hamiltonian with a Markov chain such as ours. It may also
be possible to construct a verifier using phase estimation on this
Hamiltonian.%
}

\begin{algorithm}[H]
$\hphantom{}$
\begin{enumerate}
\item \label{enu:Prepare-the-state}Take the input state $|\phi\rangle$,
append an ancilla register which holds integers $i\in\{0,...,q_{max}\}$
and prepare the state \[
|\phi^{\prime}\rangle=U\left(|\phi\rangle|0\rangle\right)\]
 using the unitary $U$ defined above.
\item \label{enu:Adjoin-two-ancilla}Adjoin an ancilla register which holds
integers $s\in\{1,...|\mathcal{S}|\}$ and initialize the ancilla
register to the state $\sum_{s}\frac{1}{\sqrt{\left|\mathcal{S}\right|}}|s\rangle$.
This produces the state \[
|\phi^{\prime}\rangle\sum_{s}\frac{1}{\sqrt{\left|\mathcal{S}\right|}}|s\rangle\]
 on three registers (one which holds grid diagrams $G$, one which
holds integers $i\in\{0,...,q_{max}\}$ and one which holds integers
$s\in\{1,...,|\mathcal{S}|\}).$
\item \label{enu:Now-apply-the}Repeat $r=\mathop\mathrm{poly}(\bar{D})$
times:

\begin{enumerate}
\item \label{enu:Apply-the-unitary}Apply the unitary $V$, where \[
V=\sum_{s}\hat{P}_{s}\otimes|s\rangle\langle s|.\]
This operator applies the permutation $\hat{P}_{s}$ to the first
two registers, conditioned on the value $s$ in the third register.
\item \label{enu:step_b}Measure the projector \[
Q=I\otimes I\otimes\left(\sum_{s,s^{\prime}}\frac{1}{\left|\mathcal{S}\right|}|s\rangle\langle s^{\prime}|\right).\]

\end{enumerate}
\item \label{Accept/reject }If you obtained the outcome 1 in each of the
$r$ repetitions of step \ref{enu:Now-apply-the}, then accept the
money. In this case apply $U^{\dagger}$ and then remove the second
and third registers. The final state of the first register is the
output quantum money state. If you did not obtain the outcome $1$
in each of the $r$ iterations in step \ref{enu:Now-apply-the}, then
reject the money.
\end{enumerate}
\caption{Markov chain verification algorithm}

\end{algorithm}

For a quantum money state $|\$_{p}\rangle$ the associated state $|\$_{p}^{\prime}\rangle\sum_{s}\frac{1}{\sqrt{\left|\mathcal{S}\right|}}|s\rangle$
prepared in steps \ref{enu:Prepare-the-state} and \ref{enu:Adjoin-two-ancilla}
is unchanged by applying the unitary $V$ in step \ref{enu:Apply-the-unitary}.
Measuring the projector $Q$ in step \ref{enu:step_b} on this state
gives $+1$ with certainty. So for good quantum money, all the measurement
outcomes obtained in step \ref{Accept/reject } are $+1$. We can
see that good quantum money passes verification. 

Now let us consider the result of applying the above procedure to
a general state $|\phi\rangle$. The first step of the algorithm is
to prepare $|\phi'\rangle=U\left(|\phi\rangle|0\rangle\right)$. If
a single iteration of the loop in step \ref{enu:Now-apply-the} results
in the outcome 1 then the final state of the first two registers is
\[
\frac{\frac{1}{\left|\mathcal{S}\right|}\sum_{s}\hat{P}_{s}|\phi^{\prime}\rangle}{\left\Vert \frac{1}{\left|\mathcal{S}\right|}\sum_{s,t}\hat{P}_{s}|\phi^{\prime}\rangle\right\Vert }.\]
This occurs with probability \begin{equation}
\left\Vert \frac{1}{\left|\mathcal{S}\right|}\sum_{s}\hat{P}_{s}|\phi^{\prime}\rangle\right\Vert ^{2}.\label{eq:prob}\end{equation}
The entire procedure repeats the loop $r$ times and the probability
of obtaining all outcomes $1$ is \[
\left\Vert \left(\frac{1}{\left|\mathcal{S}\right|}\sum_{s}\hat{P}_{s}\right)^{r}|\phi^{\prime}\rangle\right\Vert ^{2},\]
in which case the final state (on the first two registers) is \[
\frac{\left(\frac{1}{\left|\mathcal{S}\right|}\sum_{s}\hat{P}_{s}\right)^{r}|\phi^{\prime}\rangle}{\left\Vert \left(\frac{1}{\left|\mathcal{S}\right|}\sum_{s}\hat{P}_{s}\right)^{r}|\phi^{\prime}\rangle\right\Vert }.\]

To get some intuition about what states might pass even the first
iteration of this test with reasonable probability (\ref{eq:prob}),
note that the state $\frac{1}{|\mathcal{S}|}\sum_{s}\hat{P}_{s}|\phi^{\prime}\rangle$
can only have norm close to 1 if most of its terms add coherently.
In other words, most of the $\hat{P}_{s}|\phi^{\prime}\rangle$ must
be nearly the same (they are all exactly the same for good quantum
money).

Since $\frac{1}{\left|\mathcal{S}\right|}\sum_{s}\hat{P}_{s}=\hat{B}$
is our Markov matrix, the set of states that pass all of the rounds
is directly related to the mixing properties of the Markov chain---these
states correspond to eigenvectors of $\hat{B}$ with eigenvalues close
to 1.

\subsection{Security of the money scheme\label{sub:Discussion}}

We have shown that the quantum money states $|\$_{p}\rangle$ can
be efficiently generated and pass verification with high probability.
In this section we discuss why we believe this quantum money is hard
to forge. We consider four possible types of attack against our quantum
money.

First, an attacker might measure a valid quantum money state $|\$_{p}\rangle$
to learn a grid diagram with Alexander polynomial $p$ and then generate
a superposition containing that diagram that passes verification.
One such state is the (correctly weighted) superposition over grid
diagrams equivalent to the measured diagram. If an attacker could
do this, the attacker's algorithm could be used to solve grid diagram
equivalence, i.\,e.\ the problem of telling whether or not two grid
diagrams represent the same link. This is believed to be a hard problem
on average, even for quantum computers. In fact, even deciding whether
or not a grid diagram represents the unknot is conjectured to be hard.%
{}

%
{} Second, there are likely to exist grid diagrams of dimension $2\bar{D}$
(our maximum grid dimension) where no allowed grid move reduces the
dimension (this follows from the fact that every link has a minimum
dimension of grid diagrams that represent it). Because we disallow
moves which increase the dimension above $2\bar{D}$, the Markov chain
starting from one of these grid diagrams with dimension $2\bar{D}$
will only mix over a small set of diagrams with the same dimension.
Uniform superpositions over these sets will pass step \ref{enu:Apply-the-Markov}
of the verification procedure. Step \ref{enu:Measure-the-projector}
of the verification procedure is designed to reject such superpositions.

Third, if the Markov chain does not mix well, then there will be eigenstates
of the matrix $\hat{B}$ with eigenvalues near +1. For such eigenstates
there may be counterfeit quantum money states which will pass verification
with nonnegligible probability. We do not know if these states exist,
and even if they do we do not know how to make them. In fact even
the simpler question, of finding two equivalent diagrams which require
a super-polynomial number of grid moves to go from one to the other
(or proving such diagrams do not exist) seems hard.%
\footnote{Hass and Nowik \cite{hass-unknot} recently gave the first example
of a family of knots that are equivalent to the unknot and require
a quadratic number of Reidemeister moves to untangle (previous lower
bounds were only linear). %
}

Fourth, the attacker could use a valid money state $|\$_{p}\rangle$
and attempt to generate $|\$_{p}\rangle\otimes|\$_{p}\rangle$ (or
some entangled state on two registers where each register would pass
verification). Such an attack worked to forge product state quantum
money using quantum state restoration \cite{state-restoration}. However,
in that case the attack works because the product state money has
an embedded classical secret that the attacker can learn to forge
the state. In the present work, there is no obvious secret hidden
in the money or in the verification algorithm that an attacker could
use. 

The list above comprises all the lines of attack we were able to think
of.

\subsection{Why not quantum money from graphs?\label{sub:Why-not-Quantum}}

We now discuss an alternative quantum money scheme that is similar
to the one presented in this paper, but which is based on graphs rather
than knots. In practice, it is easy to find the isomorphism relating
most pairs of isomorphic graphs; this is enough to render graph based
quantum money unusable. The knot based quantum money is an analog
of this scheme (translating various concepts from graphs to knots)
but we believe it is secure since knot equivalence is believed to
be difficult on average.

In the graph based quantum money scheme, the Hilbert space has basis
vectors which encode adjacency matrices of graphs on $n$ vertices.
The mint generates the state\[
\sum_{\text{Adjacency matrices \ensuremath{A}}}|A\rangle|0\rangle.\]
The mint then computes the eigenvalues of the adjacency matrix into
the second register and measures the second register to obtain a particular
spectrum $R=\left\{ \lambda_{1},...\lambda_{n}\right\} .$ The resulting
state is \[
|\$_{R}\rangle|R\rangle=\sum_{A\text{ with spectrum }R}|A\rangle|R\rangle.\]
 The quantum money is the state $|\$_{R}\rangle$ and the serial number
encodes the spectrum $R$. The verification procedure is based on
a classical Markov chain that, starting from a given adjacency matrix
$A$, mixes to the uniform distribution over adjacency matrices $A^{\prime}$
that represent graphs isomorphic to $A$. 

Given two adjacency matrices $A_{0}$ and $\sigma A_{0}\sigma^{-1}$
that are related by the permutation $\sigma\in S_{n}$, the forger
can usually efficiently find the permutation $\sigma$ (we assume
for simplicity that $A_{0}$ has no automorphisms so this permutation
is unique). We now show how this allows a counterfeiter to forge the
graph based quantum money. The forger first measures the state $|\$_{R}\rangle$
in the computational basis, obtaining a particular adjacency matrix
$A$ with the spectrum $R$. Starting from this state, the forger
adjoins two additional registers (one which holds permutations and
one which holds adjacency matrices) and, applying a unitary transformation,
the forger can produce the state

\[
\sum_{\pi\in S_{n}}|A\rangle|\pi\rangle|\pi A\pi^{-1}\rangle.\]
 Now the forger can use the procedure which solves graph isomorphism
to uncompute the permutation which is held in the second register,
producing the state \[
|A\rangle|0\rangle\sum_{\pi\in S_{n}}|\pi A\pi^{-1}\rangle.\]
 The state of the third register will pass verification. The forger
can repeat this procedure using the same adjacency matrix $A$ to
produce many copies of this state which pass the verification as quantum
money.

Return to the knot based quantum money. Given two grid diagrams $G_{1}$
and $G_{2}$ which represent the same link, we believe it is hard
(on average) to find a sequence of grid moves which transform $G_{1}$
into $G_{2}$. Even given an oracle for the decision problem of determining
whether two links are equivalent, we do not know how to find a sequence
of Reidemeister moves relating the links. We hope this discussion
has motivated some of the choices we have made for our knot based
quantum money.

\section{Conclusions\label{sec:Conclusions}}

Forge it if you can.

\section{Acknowledgments}

We thank Louis Kauffman, Haynes Miller and Tom Mrowka for valuable
discussions on knot theory. This work was supported in part by funds
provided by the U.S. Department of Energy under cooperative research
agreement DE-FG02-94ER40818, the W. M. Keck Foundation Center for
Extreme Quantum Information Theory, the U.S. Army Research Laboratory's
Army Research Office through grant number W911NF-09-1-0438, the National
Science Foundation through grant number CCF-0829421, the DoD through
the NDSEG Fellowship Program, the Natural Sciences and Engineering
Research Council of Canada, and Microsoft Research.

\bibliographystyle{plain}
\addcontentsline{toc}{section}{\refname}\bibliography{knot-bib}

\appendix

\appendix

\section{Details of the Markov chain on planar grid diagrams}

The Markov chain in section \ref{sub:A-Markov-Chain} acts on the
configuration space of pairs $(G,i)$ where $1\leq i\leq q(d(G))$.
Here we describe an algorithm to implement this Markov chain. For
each update of the state of the Markov chain, we first choose several
uniformly random parameters: \begin{align*}
j: & \quad\text{an integer from 1 to 8}\\
w: & \quad\text{an integer from 0 to \ensuremath{q_{max}^{2}}}\\
x,y: & \quad\text{integers from 0 to \ensuremath{2\bar{D}-1}}\\
k: & \quad\text{an integer from 0 to 3 }\end{align*}
where $q_{max}$ is the maximum value attained by the function $q(d)$
for $d\in\{2,...,2\bar{D}\}.$ We then update the configuration $(G,i)$
in a manner that depends on the values of these variables, where $j$
picks the type of the update being performed, and the other variables
are used as parameters:
\begin{itemize}
\item If $j=1$, set \[
i\leftarrow\left(i+w\right)\text{ mod \ensuremath{q(d(G))}}\]
Leave $G$ unchanged. This is the only move which changes $i$, and
it is used to mix between different labels.
\item If $j=2$, cyclically permute the columns of $G$ by moving each column
to the right by one. Leave $i$ unchanged.
\item If $j=3$, cyclically permute columns of $G$ by moving each column
to the left by one. Leave $i$ unchanged.
\item If $j=4$, cyclically permute the rows of $G$ by moving each row
up by one. Leave $i$ unchanged.
\item If $j=5$, cyclically permute the rows of $G$ by moving each row
down by one. Leave $i$ unchanged.
\item If $j=6$ and $x+1<d\left(G\right)$, then check whether transposing
columns $x$ and $x+1$ is a legal move (as defined in section~\ref{sub:Grid-diagrams}).
If so, transpose them; otherwise, do nothing. Leave $i$ unchanged.
\item If $j=7$ and $y+1<d\left(G\right)$, then check whether transposing
rows $y$ and $y+1$ is a legal move (as defined in section~\ref{sub:Grid-diagrams}).
If so, transpose them; otherwise, do nothing. Leave $i$ unchanged.
\item If $j=8$, $k=0$, $x,y<d\left(G\right)$, and there is a marker (X
or O) at position $\left(x,y\right)$, then consider stabilizing by
adding markers forming an L to the upper right of position $(x,y)$.
In this case, construct $G^{\prime}$ from $G$ by adding a new column
to the right of $x$ and a new row above $y$, deleting the original
marker, adding markers of the same type at $\left(x+1,y\right)$ and
$\left(x,y+1\right)$, and adding a marker of the opposite type at
$\left(x+1,y+1\right)$. Then if $i\leq q(d(G^{\prime})),$ set $(G,i)\leftarrow(G^{\prime},i)$.
If not, leave the configuration unchanged.
\item If $j=8$, $k=0,$ $x,y<d\left(G\right)-1$, there is no marker at
$(x,y)$, and there are markers at positions $\left(x+1,y+1\right)$,
$\left(x+1,y\right)$, $\left(x,y+1\right)$, then consider destabilizing
by removing markers forming an L to the upper right of position $\left(x,y\right)$.
In this case, construct $G^{\prime}$ from $G$ by removing column
$x+1$ and row $y+1$ (thus removing those three markers) and adding
a new marker of the appropriate type at $\left(x,y\right)$ (measured
after deletion of the row and column). If $i\leq q(d(G^{\prime})),$
set $(G,i)\leftarrow(G^{\prime},i)$. If not, leave the configuration
unchanged.
\item If $j=8$, and $k\in{1,2,3}$, rotate the grid diagram by $90k$ degrees
counterclockwise and then apply the update rule corresponding to $j=8,k=0$
with the same $(x,y)$. After applying the update, rotate the diagram
back by the same amount.
\end{itemize}
The parameter $j$ determines the type of move. The first move ($j=1)$
is used to permute between different labels of the same graph. It's
the only move which changes the labels. Moves 2--5 are cyclic permutations
of rows and columns, while moves 6 and 7 are transpositions. Finally,
$j=8$ stabilizes or destabilizes an L in a direction (\tikz[x=1ex,y=1ex]\draw (1,0) -- (1,1) -- (0,1);
\tikz[x=1ex,y=1ex]\draw (1,1) -- (1,0) -- (0,0); \tikz[x=1ex,y=1ex]\draw (0,1) -- (0,0) -- (1,0);
or \tikz[x=1ex,y=1ex]\draw (0,0) -- (0,1) -- (1,1);) depending on
$k$ (0, 1, 2, or 3 respectively). In all the moves $x$ is treated
as column index (which describes where to apply the move), and $y$
is treated as a row index.

For fixed $\left(j,w,x,y,k\right)$, the update of the state is an
easily computable, invertible function (in other words it is a permutation
of the configuration space). This is easy to see for any of the moves
with $j\in\{1,...,7\}$. One can check that each move with $j=8$
is its own inverse. This justifies our assertion that the Markov matrix
$B$ can be written in the form of equation \ref{eq:B_matrix}.

In the notation of Section~\ref{sub:Verifying-quantum-money}, $s=\left(j,w,x,y,k\right)$
and \[
\mathcal{S}=\{1,\ldots,8\}\times\{0,\ldots,q_{\text{max}}^{2}\}\times\{0,\ldots,2\bar{D}-1\}\times\{0,\ldots,2\bar{D}-1\}\times\{0,1,2,3\}.\]
 
\end{document}